# Silicene Spintronics

Yangyang Wang (王洋洋),[1,3 †] Ruge Quhe (屈贺如歌),[1,4 †] Dapeng Yu (俞大鹏),[1,2] Ju Li (李巨),[3] and Jing Lu (吕劲)[1,2 *]

[1]State Key Laboratory for Mesoscopic Physics and Department of Physics, Peking University, Beijing 100871, P. R. China

[2]Collaborative Innovation Center of Quantum Matter, Beijing 100871, P. R. China

[3]Department of Nuclear Science and Engineering and Department of Materials Science and Engineering, Massachusetts Institute of Technology, Cambridge, Massachusetts 02139, USA

[4]Academy for Advanced Interdisciplinary Studies, Peking University, Beijing 100871, P. R. China

*Corresponding author: jinglu@pku.edu.cn

†These authors contributed equally to this work.

## Abstract

Spintronics involves the study of active control and manipulation of spin degrees of freedom in solid-state systems. The fascinating spin-resolved properties of graphene motivate numerous researchers into the studies of spintronics in graphene and other two-dimensional (2D) materials. Silicene, silicon analog of graphene, is considered as a promising material for spintronics. Here, we present a review on the theoretical advances about the spin-dependent properties including the electric field and exchange field tunable topological properties of silicene and the corresponding spintronic device simulations.

# Introduction

Spintronics, or spin electronics, aims to exploit spin degrees of freedom instead of or in addition to charge degrees of freedom for information storage and logic devices. Compared with conventional semiconductor devices, spintronic devices have the potential advantages of nonvolatility, increased data processing speed, decreased electric power consumption, and increased integration densities. Currently, the major challenge of spintronics is the difficulty in generating, controlling and detecting spin-polarized current. Graphene, with a high electronic mobility of 200,000 cm$^2$/(V·s),[1, 2] and a long spin diffusion length of several micrometers,[3-5] is quite attractive for application in spintronics. Since its first isolation in 2004, a variety of graphene-based spintronic devices have been proposed in the past decade,[6, 7] such as spin valve,[8] large magnetoresistance (MR) devices,[9] and graphene nanoribbon (GNR)-based giant MR device, spin-filter and spin transistors.[10-14]

Silicene, silicon counterpart of graphene, is predicted to be another Dirac material and has been successfully fabricated via epitaxial growth on the Ag(111),[15-19] ZrB$_2$(0001),[20] Ir(111),[21] and MoS$_2$[22] surfaces recently. Silicene filed effect transistor (FET) operating at room temperature has been fabricated very recently,[23] corroborating theoretical expectations regarding its ambipolar Dirac charge transport. Compared with graphene, silicon has a longer spin-diffusion time ($\tau_s$ = 1 ns at 85 K and 500 ns at 60 K $vs$ $\tau_s$ = 0.1 ns at 300 K), a longer spin coherence length ($l_s$ = 10, 350, 2000 μm $vs$ $l_s$ = 1.5 and 2 μm at room temperature), and a much larger spin–orbit coupling (SOC) gap (1.55 meV $vs$ 10$^{-3}$ meV). [3, 24-29] Therefore, silicene appears more suitable for spintronic applications and could fit more easily into industrial silicon-based circuits than graphene.

In spintronics, half-metallic materials, with metallic nature for one spin and insulating or semiconducting nature for the other spin, are highly desired because they could provide 100% spin polarized current. Due to its peculiar edge states, zigzag silicene nanoribbons (ZSiNRs) provide a good platform for the realization of half-metallicity by transverse electric filed[30] and asymmetric edge modification.[31-34] Spin-filter and spin FET are designed based on the half-metallicity in ZSiNRs. Besides, a giant MR is obtained in ZSiNRs either by utilizing a switchover between different magnetic states[35] or symmetry-dependent transport property.[36]

Half-metallicity can also be achieved in semihydrogenated silicene by applying an out of plane gate voltage. By adjusting an electric field and/or an exchange field, quantum spin Hall effects (QSHE), quantum anomalous Hall effects (QAHE), valley-polarized QAHE, and quantum valley Hall effects (QVHE) can be expected in pristine silicene, which provide many potential applications for spintronic devices. This review is organized into the corresponding three sections as follows.

## 1. Spintronics in zigzag silicene nanoribbons

ZSiNRs have drawn a lot of attention due to their remarkable application potential in spintronics. The ground state of ZSiNRs has the two edges antiferromagnetically (AFM) coupled, which is slightly lower in energy than the ferromagnetical (FM) state as a result of edge magnetic states coupling.[30] Due to the spin degeneracy of the AFM state, spin-polarized transport can't be realized in pristine ZSiNRs (Fig. 1). Therefore, various approaches were utilized to break the spin degeneracy in ZSiNRs, such as applying external electric field[30] and edge modification, [31-34] just like in the zigzag GNRs.[11, 14]

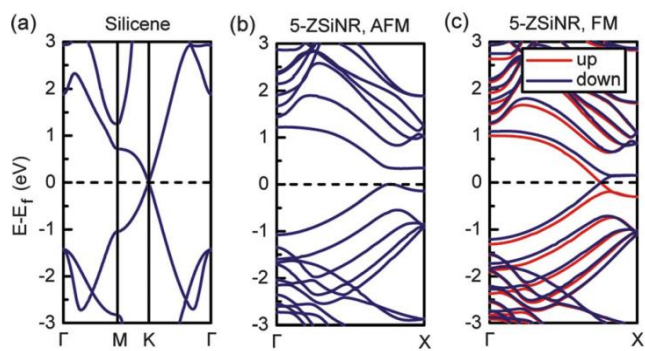

**Figure 1.** Band structures of (a) silicene, (b) the AFM 5-ZSiNR, and (c) the FM 5-ZSiNR. Reproduced with permission from Ref. [35]. Copyright 2012 Royal Society of Chemistry.

### 1.1 Half-Metallicity by a Transverse Electric Field

The half-metallicity could be induced by an in-plane homogenous electrical field in the hydrogen-terminated ZSiNRs.[30] The zero-field $α$-spin and $β$-spin orbitals in the conduction and valence bands are localized at the opposite edges of the nanoribbon, and in the same edge the spin orientations in the conduction and valence bands are opposite (Fig. 2c). As the transverse electric field $E_{ext}$ increases, the spin degeneracy of the conduction and valence bands is lifted. The band gap of $\beta$-spin state decreases and finally closes under $E_{ext}$ = 0.25 V/Å, while that of $\alpha$-spin state increases slightly relative to the zero-field value. Therefore, the transverse electrical field-induced half-metallicity in ZSiNRs is well established (Fig. 2b).

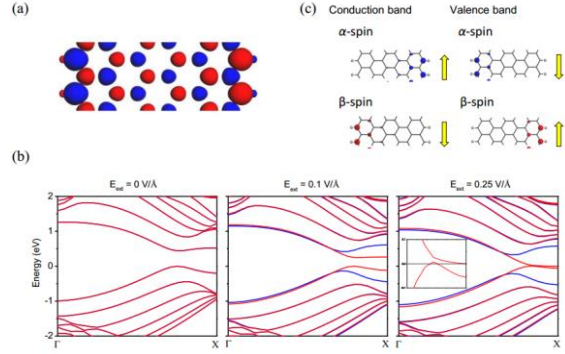

**Figure 2**. Electronic properties of the 6-ZSiNR in the ground state. (a) Spatial spin density distribution. The isovalue is 0.006 a.u. (b) Spin-resolved band structures under $E_{ext}$ = 0, 0.1, and 0.25 V/Å, respectively. Inset: the band structure with $E_{ext}$ = 0.25 V/Å in the range of $|E| <$ 0.1 eV and $0.7\pi/a \leq k \leq \pi/a$ (the horizontal line is $E_f$). The valence top or $E_f$ is set to zero. (c) $\alpha$-spin and $\beta$-spin orbitals of the conduction and valence band, shown as the square of the absolute value of the wavefunction summed over all $k$-points. The isovalue is 0.275 a.u. The yellow arrow represents the energy shift direction of the spin states under a transverse electrical filed. Blue and red denote $\alpha$-spin and $\beta$-spin, respectively. Reproduced with permission from Ref. [30]. Copyright 2012 World Scientific Publishing Company.

**1.2 Half-Metallicity by Asymmetric Edge Hydrogenation**

Asymmetric edge modification is another way to transform ZSiNRs to half-metals.[31-34] One of these interesting methods is asymmetric edge hydrogenation ($H_2$–ZSiNR–H).[31, 33, 34] Several theoretical works have demonstrated that the ground state of $H_2$–ZSiNR–H is a FM semiconductor. Around the Fermi level ($E_f$), the states are completely spin-polarized with opposite spin orientations, declaring a bipolar magnetic feature in the asymmetric ZSiNRs (Fig. 3).[31] This special distribution will bring interesting magnetic behaviors for doped $H_2$–ZSiNR–H. In doped systems, $E_f$ will shift into the valence or conduction bands, and $H_2$–ZSiNR–H becomes a half metal.

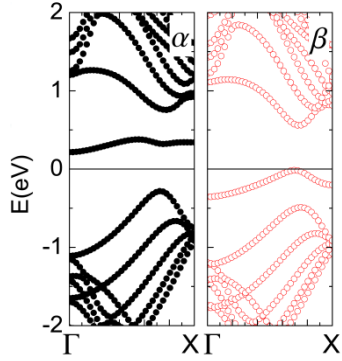

**Figure 3.** The spin-polarized band structures for the asymmetric 6-ZSiNRs. Reproduced with permission from Ref.[31]. Copyright 2013 American Physical Society.

**1.3 Spin-filter**

A ZSiNR spin-filter under a transverse electrical field is illustrated in Fig. 4a.[30] Transportation properties are simulated by using the density functional theory (DFT) coupled with nonequilibrium Green's function (NEGF) formalism. In the spin-resolved transmission spectrum of this device under $E_{gate}$ = 0.3 V/Å, there is a large peak for α-spin around $E_f$ and a clear gap for β-spin (Fig. 4b). Consistently，the spin-resolved transmission eigenstates at $E_f$ and the Γ point in $k$-space in Fig. 4c also reflect this highly spin polarization. The spin-filter efficiency, defined as SFE= $(I_α - I_β)/( I_α + I_β)$, where $I_α$ and $I_β$ represent the α-spin and β-spin current densities, respectively. The changes of $T(E_f)$ and SFE as a function of $E_{gate}$ are shown in Fig. 4d. The $T(E_f)$ for the two spins are symmetric about $E_{gate}$ = 0. Even at small $E_{gate}$ = -0.05 or 0.05 V/Å, the $T(E_f)$ between the two spins has a difference with SFE = 62.3%. The difference becomes more and more significant with the increasing $E_{gate}$, a behavior consistent of the electrical field-induced change in the band gap of ZSiNRs. SFE is nearly saturated (99%) from $|E_{gate}|$ > 0.2 V/Å. Therefore, the dual-gated finite ZSiNR can serve as a nearly perfect spin-filter, with sign switchable by altering the electric field direction. Similarly, a perfect spin filtering effect can be expected in a doped half metal $H_2$-ZSiNR-H. The study shows that a SFE of 100% is achieved with the unchanged spin states in a very large bias region from −0.5 V to 0.5 V without the magnetic field being applied (Fig. 5).[34]

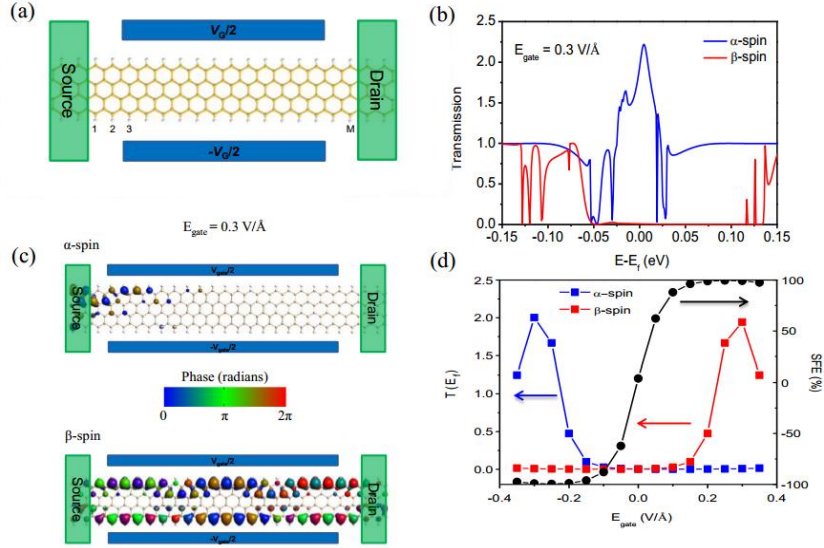

**Figure 4.** Spin-filter based on the 4-ZSiNR. (a) Schematic model with one pair of gate electrodes on the two sides. (b) Spin-resolved transport spectrum under $E_{gate}$ = 0.3 V/Å. (c) Spin-resolved transmission eigenstate at $E_f$ and the $\Gamma$ point in *k*-space under $E_{gate}$ = 0.3 V/Å. The isovalue is 1.0 a.u. (d) Spin-resolved transmission coefficient at $E_f$ *k*-space and spin filtration efficiency as a function of $E_{gate}$. Reproduced with permission from Ref. [30]. Copyright 2012 World Scientific Publishing Company.

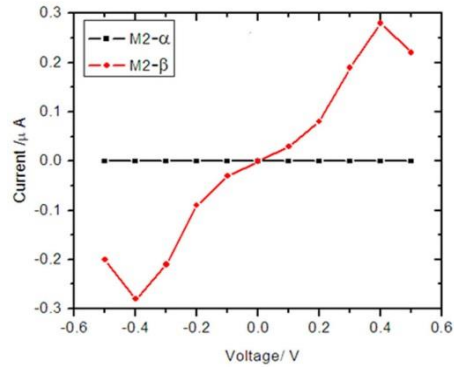

**Figure 5.** The *I* – *V* curves for the *α*-spin and the *β*-spin states at various biases. Reproduced with permission from Ref. [34]. Copyright 2013 Nature Publishing Group.

**1.4 Spin FET**

Besides the proposed spin-filter, ZSiNR may also serve as the channel of an effective spin FET with quadruple-gate, as illustrated in Fig. 6.[30] The electrical field of the left pair of electrodes ($E_{LG}$) is fixed and the on-off switch is achieved by modulating the right one ($E_{RG}$).

When $E_{RG} = E_{LG}$, the left and right parts of the nanoribbon allow the same spin to transport along the edges. This device degenerates into a spin-filter with $E_{gate} = E_{RG} = E_{LG}$. When the direction of the electrical field of the right pair of electrodes is reversed, the sign of the allowed travelling spin in the right part of the nanoribbon is reversed and contrary to that in the left part, resulting in a possible blockade of the transmission of both spins. As a result, the current of this device is expected to be forbidden in this case. Therefore, through altering $E_{RG}$ and thus altering the spin state, the quadruple-gated device can operate as a spin FET. The maximal on/off conductance ratio of the present device is 18. The low on/off ratio is ascribed to the quiet short channel (38.3 Å) controlled by each pair of electrodes in this simulation limited by the computational resource, which gives rise to a certain amount of leakage current on the off-state due to tunneling effect. If the channel length is increased, a higher on/off ratio is expected because the leakage of the two spin currents will both be reduced.

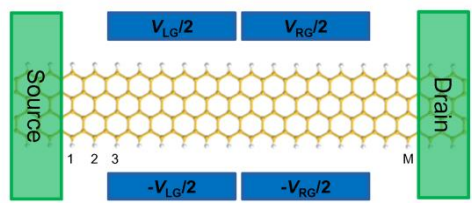

**Figure 6**. Schematic model of the quadruple-gated spin FET based on the 4-ZSiNR. Reproduced with permission from Ref. [30]. Copyright 2012 World Scientific Publishing Company.

**1.5. Giant Magnetoresistance**

The AFM state of ZSiNRs is semiconducting, while the FM state is metallic. If applying a proper magnetic field, ZSiNRs can switch between the AFM and the FM configurations (Figs. 7a and 7b), and consequently a large MR can be obtained because of a large current difference between the semiconducting AFM and the metallic FM configurations (Fig. 7c). Usually the optimistic definition is adopted to calculate MR, which is defined as $MR(V_{bias}) = (I_{FM} - I_{AFM})/I_{AFM}$, where $V_{bias}$ is the applied bias voltage, $I_{FM}$ ($I_{AFM}$) is the total current of the FM (AFM) configuration. As shown in Fig. 7d, the MRs of different ZSiNRs at the same bias voltage drop generally with the increasing ribbon width, except 4-ZSiNR. In the examined bias range from 0.05 to 0.5 V, the maximum MRs of the ZSiNRs are 261–1960%.

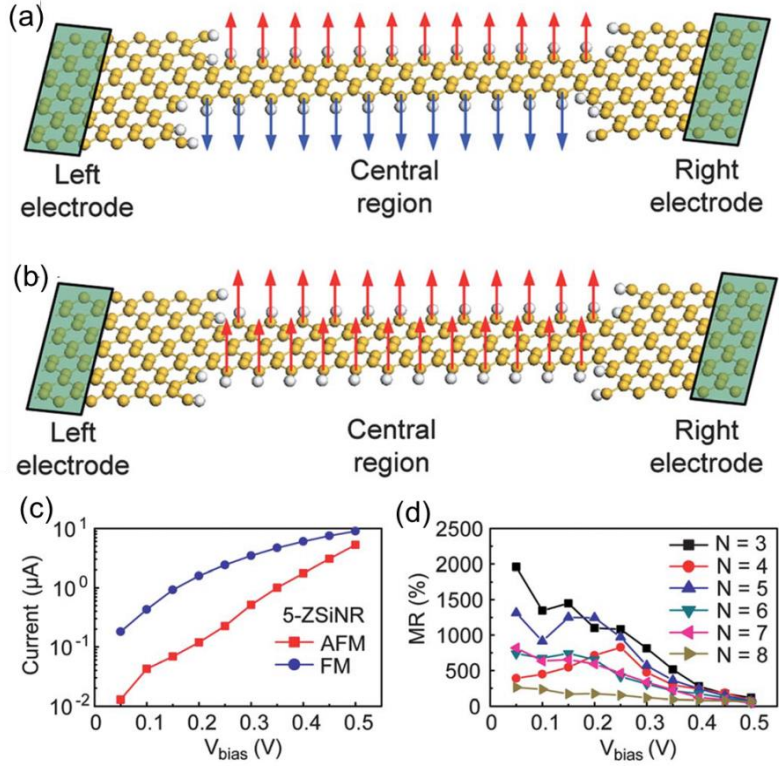

**Figure 7.** (a) and (b) Schematic models of a H-passivated ZSiNR connected to two semi-infinite silicene structures. (a) The 3-ZSiNR in the AFM configuration. (b) The 4-ZSiNR in the FM configuration. Applying or removing magnetic field adjusts the ferromagnetic coupling of the two edges. The yellow (gray) balls denote silicon (hydrogen) atoms. The arrows represent spin directions on the edges. (c) $I$–$V_{bias}$ characteristics of the 5-ZSiNR at the AFM and FM configurations. (d) MRs of different-width ZSiNRs as a function of bias voltage. Reproduced with permission from Ref. [35]. Copyright 2012 Royal Society of Chemistry.

The critical magnetic field $B^*$ required to switch a ZSiNR between the AFM and FM configurations can be estimated from the relation:

$$B^* = \frac{\Delta}{g\mu_B M_T} \quad (1)$$

where $\Delta$ is the energy difference between the AFM and FM configurations, $g = 2$ is the Landauer factor of silicene, $\mu_B = 0.058$ meV/T is the Bohr magneton, and $M_T$ is the total spin on edge atoms. The resulting $B^*$ of the 5-ZSiNR is 8.6 T, which is easily attainable in the laboratory.

Apart from applying a magnetic field, peculiar symmetry-dependent transport properties of FM ZSiNRs can also be utilized to achieve a giant MR.[36] Due to the existing of a two-fold axis, the $\pi$ and $\pi^*$ wavefunctions in even-$N$ ZSiNRs have opposite parity with respect to the $c2$ operation. The transmission of electron from the $\pi$ band of the left electrode to the $\pi^*$ band of the right electrode is forbidden, leading to a conductance gap near $E_f$. On the other hand, the transmission is allowed in odd-$N$ ZSiNRs due to the the $\pi$ and $\pi^*$ wavefunctions have no definite parity.

Two spin configurations P and AP are considered in a two-probe system of even-$N$ ZSiNRs (Figs. 8b-8c). Both the left and right electrodes are spin-up polarized in P configuration, while in AP configuration the two electrodes have antiparallel spin polarization direction. Fig. 8d shows the $I$-$V$ relationship for different spin components. The magnitude of MR can be obtained according to the definition MR = $(I_P - I_{AP})/I_{AP}$, where $I_P$ and $I_{AP}$ are currents in P and AP configurations, respectively. The spin-up, spin-down, and total MRs are all in the order of 1 000 000% (Fig. 8e). The appearance of MR effect can be understood by Figs. 8f-8g. For P configuration (Fig. 8f), near $E_f$ the spin-up $\pi$ bands, as well as the spin-down $\pi^*$ bands, of the two electrodes overlap, and the transmission is allowed. So near the Fermi level, the conductance is about 1 $G_0$ for both spin components, and the current increases with increasing bias. The situation is different in AP configuration. As shown in Fig. 8g, near $E_f$ the spin-up $\pi^*$ (spin-down $\pi$) band of left electrode only overlaps with the spin-up $\pi$ (spin-down $\pi^*$) band of the right electrode. As discussed above, the $\pi$ and $\pi^*$ bands have opposite parity, thus the transmission between them is forbidden. As a result, a conductance gap appears around $E_f$ for both spin components, and the corresponding current is almost zero.

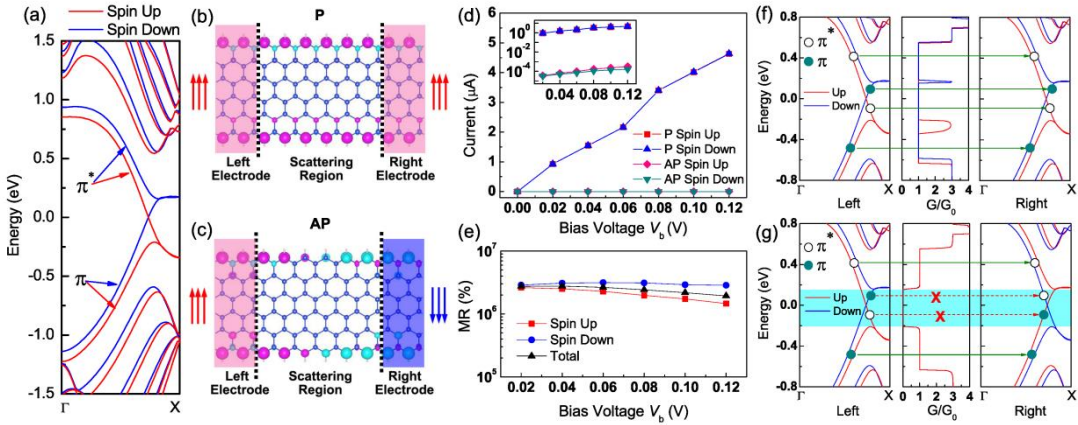

**Figure 8.** (a) Band structure of FM 6-ZSiNR. The spin up and spin down components are presented in red and blue, respectively. (b) and (c) Spin density for 6-ZSiNR with P and AP configurations under zero bias. The two electrodes have parallel spin configuration in the P

case but antiparallel spin configuration in the AP case. Pink and blue surfaces denote the spin up and spin down components, respectively. The isosurface corresponds to 0.01 $e$/Å$^3$. (d) The *I-V* curves for the P and AP configurations. The inset is semi-logarithmic scale plot. (e) The spin up, spin down, and total magnetoresistance on semi-logarithmic scale. (f) and (g) The band structures for left and right electrodes, and the transmission spectrum for P (f) and AP (g) configurations under zero bias. Solid and open circles denote $\pi$ and $\pi^*$ states. Solid arrows indicate allowed transmissions, and dashed arrows indicate forbidden transmissions. $G_0$ equals to $e^2/h$. Reproduced with permission from Ref. [36]. Copyright 2012 American Institute of Physics.

## 2. Spintronics in semihydrogenated functionalized silicene

The semihydrogenated silicene (H@Silicene) shown in Figs. 9a-9b is dynamically stable, demonstrated by molecular dynamics and phonon mode dispersion calculations.[37] According to DFT calculations, H@Silicene is a FM semiconductor with a band gap of 0.93 eV (Fig. 9c), and both electronic and spintronic applications can be expected. Regarding the spin-polarized band structures, the Fermi level can be shifted into the valence or conduction bands by applying an out-of-plane gate voltage. Therefore a half metal behavior could be expected in the H@Silicene.

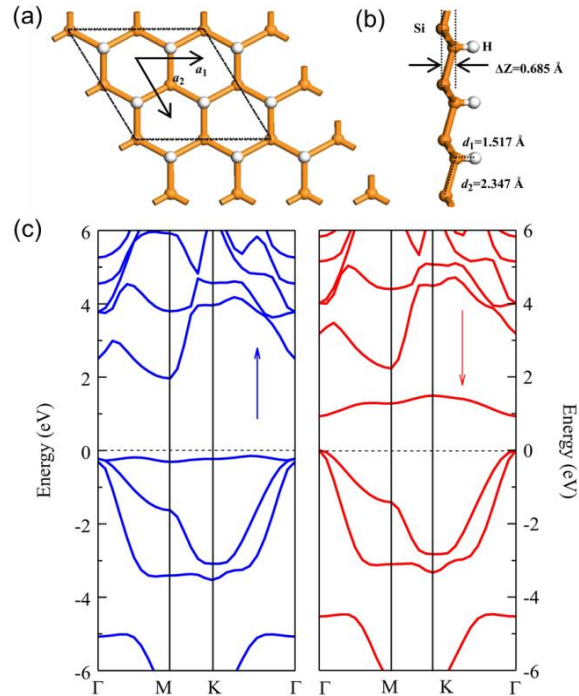

**Figure 9.** The optimized geometric atomic configurations and the structural parameters of

H@Silicene. (a) Top view with the rhombus marked in black shows the supercell. The Brava is lattice vectors of unit cell are given with $a_1 = a_2 = a = 3.899$ Å. (b) The side view. The average bond length $d_1$ (Å) between Si and H atoms, $d_2$ (Å) between Si and Si atoms, and buckled height ΔZ (Å) between Si and Si layers. The yellow and white balls stand for Si and H atoms, respectively. (c) Band structures of H@Silicene. The arrow denotes the spin-polarized direction. The top of the valence band is set to zero. Reproduced with permission from Ref. [37]. Copyright 2013 Elsevier B.V.

A single-gated H@Silicene device as shown in Fig. 10a is simulated by using the DFT-NEGF methods. As discussed above, by changing the gate voltage to shift $E_f$ into the valence or conduction bands, a half metal behavior could be expected in the H@Silicene device. Therefore, spin-filters and spin switches can be achieved. As shown in Fig. 10b, when $V_g = 0$ V, the transmission probability nearly vanishes, responsible for the off-state for spin up and spin down. $E_f$ shifts upward as $V_g$ increases (Fig. 10c). When $V_g = 1.9$ V (Fig. 10d), $E_f$ is located in the middle of a subband of spin down and an effective on-state for spin down is achieved. With the increase of the gate voltage, the current density of the spin down increases significantly, but that of the spin up value increases slightly, as shown in Fig. 10e. As a result, the SFE increases with the increasing gate voltage and reaches 100% at $V_g = 1.9$ V.

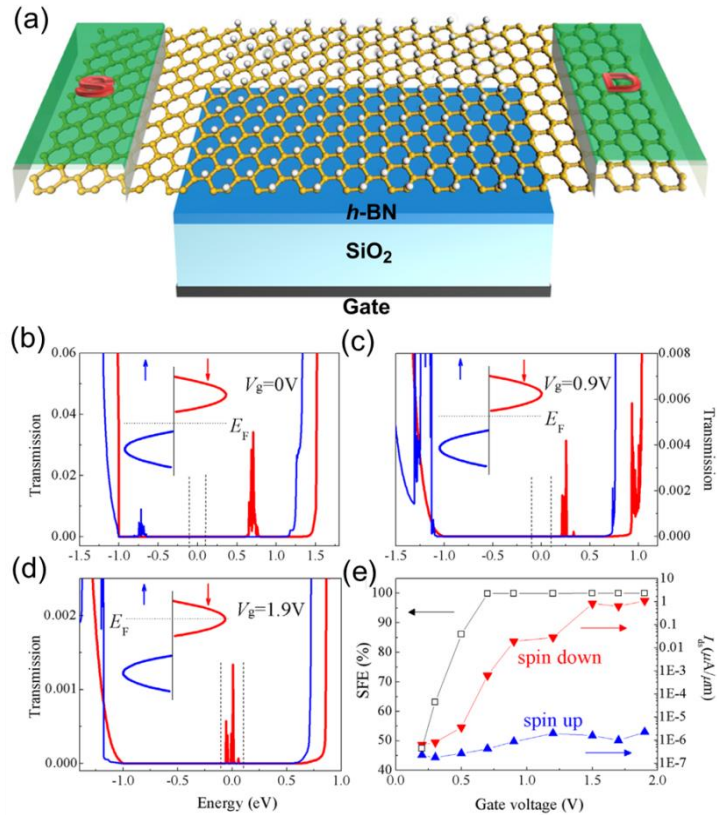

**Figure 10.** (a) Schematic of two-probe model of H@Silicene sheet spin-filter device with SiO$_2$ dielectric and h-BN buffer layer. (b), (c) and (d) Spin-polarized transmission spectra with $V_g$ = 0, 0.9 and 1.9 V, respectively. The bias voltage is fixed at $V_{bias}$ = 0.2 V. Red (blue) line stands for the spin down (up). The vertical dashed-lines denote the bias voltage window. The insets are the schematic of the Fermi level shift with the gate voltage. (e) Spin-filter efficiency and spin-resolved current as a function of the gate voltage. Reproduced with permission from Ref. [37]. Copyright 2013 Elsevier B.V.

## 3. Tunable quantum topological states in silicene and corresponding spintronic applications

The low-energy structure of silicene is described by Dirac electrons with relatively larger spin-orbit interaction. The phase diagram of silicene as a function of electric field and exchange field has been investigated by Ezawa.[38] A QAH insulator, valley polarized metal (VPM), QSH insulator, and band insulator (BI) appear. They are characterized by the Chern number and spin-Chern number and/or by the edge modes of a nanoribbon. In this work, the silicene system is described by the four-band second-nearest-neighbor tight binding model:[38]

$$H = -t \sum_{<ij>\alpha} c_{i\alpha}^+ c_{j\alpha} + i\frac{\lambda_{SO}}{3\sqrt{3}} \sum_{<<ij>>\alpha\beta} v_{ij} c_{i\alpha}^+ \sigma_{\alpha\beta}^z c_{j\beta} + i\lambda_{R1}(E_z) \sum_{<ij>\alpha\beta} c_{i\alpha}^+ (\boldsymbol{\sigma} \times \hat{\mathbf{d}}_{ij})_{\alpha\beta}^z c_{j\beta}$$
$$-i\frac{2}{3}\lambda_{R2} \sum_{<<ij>>\alpha\beta} \mu_i c_{i\alpha}^+ (\boldsymbol{\sigma} \times \hat{\mathbf{d}}_{ij})_{\alpha\beta}^z c_{j\beta} - l \sum_{i\alpha} \mu_i E_z c_{i\alpha}^+ c_{i\alpha} + M \sum_{i\alpha} c_{i\alpha}^+ \sigma_z c_{i\alpha}$$
(2)

where $c_{i\alpha}^+$ creates an electron with spin polarization $\alpha$ at site $i$, and $<i,j>/<<i,j>>$ runs over all the nearest or next nearest hopping sites. The first term is the usual nearest-neighbor hoping with the transfer energy $t$ = 1.6 eV. The second term is the effective SOC with $\lambda_{SO}$ = 3.9 meV, where $\boldsymbol{\sigma} = (\sigma_x, \sigma_y, \sigma_z)$ is the Pauli matrix of spin, with $v_{ij}$ =+1 if the next-nearest-neighboring hopping is anticlockwise and $v_{ij}$ = -1 if it is clockwise with respect to the positive $z$ axis. The third term is the first Rashba SOC associated with the nearest neighbor hopping, induced by the external electric field (Namely, extrinsic Rashba SOC). It satisfies $\lambda_{R1}(E_z = 0) = 0$, where $E_z$ represents the uniform electric field applied perpendicular to the silicene sheet. The forth term is the second Rashba SOC (Namely, intrinsic Rashba SOC) with $\lambda_{R2} = 0.7$ meV associated with the next-nearest neighbor hopping term, where $\mu_i = \pm 1$ for the A (B) site, and $\hat{\mathbf{d}}_{ij} = \mathbf{d}_{ij}/|\mathbf{d}_{ij}|$ with the vector $\mathbf{d}_{ij}$ connecting two sites $i$ and $j$ in the same sublattice. The fifth term is the staggered sublattice potential term, where $2l$ is the buckling distance with $l = 0.23$ Å. The sixth term is the exchange magnetization: Exchange field $M$ may come from proximity coupling to a ferromagnet such as depositing transition metal atoms to the silicene surface or depositing silicene on a ferromagnetic insulating substrate.

The phase diagram in the $E_z$-$M$ plane is provided in Fig. 11. The topological quantum numbers are Chern number $C$ and spin Chern number $C_s$. When $E_z = 0$ and $M = 0$, considering only the SOC effect, silicene is a QSH insulator ( $(C, C_s) = (0,1)$ ) with a small SOC gap of 1.55 meV,[28] much higher than that of graphene with an order of $10^{-3}$ meV,[29] as shown in Fig. 12 (QSH1) qualitatively. Therefore, it is possible to observe QSH state in experimentally accessible temperature for silicene. Along the $E_z$ axis in the phase diagram, a topological phase transition occurs from the topological insulator to trivial insulator.[39-41] As shown in the last row of Fig. 12, the spin degrees of freedom in the band structure are degenerate in QSH1, as a consequence of both time reversal symmetry and inversion symmetry (IS). By applying a perpendicular electric field $E_z$, a spin splitting caused by IS breaking is obtained.[39] At the $K$ point, the energy gap decreases linearly as $E_z$ increases until a critical field $E_c = \pm \frac{2\lambda_{SO}}{l}[\frac{\sqrt{1+(\alpha/l)^2}-1}{(\alpha/l)^2}]$ is reached, where $\lambda_{R1}(E_z) = \alpha E_z$ with $\alpha = 10^{-3}$ Å. Note that $(\alpha/l)^2 = 10^{-4}$ is negligible, $E_c \approx \pm \frac{\lambda_{SO}}{l} = 17$ meV/Å and $\lambda_{R1}$ is of the order of only 10 $\mu$eV at this critical electric field. With $E$ approaching the critical field $E_c$, the gap shrinks to zero, forming a Dirac-like cone near each valley (Spin VPM, *i.e.* SVPM). In sharp contrast to the Dirac cone in graphene, where each cone is spin degenerate, here the spin is fully polarized along the $z$ direction at each valley. The spin-polarization profile around the $K'$ point is opposite to that of the $K$ point as required by time reversal symmetry. Therefore, in the critical phase (SVPM), the physics of the system is dominated by two nearly fully spin-polarized (with opposite polarizations) Dirac-like cones at the $K$ and $K'$ points. For $E_z > E_c$, the energy gap reopens at each valley and drives the QSH phase into the topologically trivial band insulating phase with $(C,C_s) = (0,0)$.

Along the $M$ axis, due to $E_z = 0$, the first Rashba interaction vanishes. The band profiles are depicted in the first column of Fig. 12. As $M$ increases, due to the breaking of time reversal symmetry, the spin splitting occurs. The spin-up bands are lifted up while the spin-down bands are pushed down. When $|M| \leq \lambda_{SO}(1 + a^2\lambda_{R2}^2/\eta^2 v_F^2)$, there still exists a band gap of $\Delta = |M - s\lambda_{SO}|$ (QSH2). The band gap closes as $M$ reaches $s\lambda_{SO}$ (M), where $s$ is the spin chirality with $s = \pm 1$. If assuming $\lambda_{R2} = 0$, then as $|M|$ increases further, the two Dirac cones with opposite spins will cross each other around each $K_\mu$ point. Actually, the Rashba interaction ($\lambda_{R2} \neq 0$) mixes the two spin bands, creates the anticrossing, and opens a gap to form the QAH insulating state with $(C,C_s) = (2,0)$ or $(-2,0)$, depending on the direction of $M$.

When $ME_z \neq 0$, in the regions where the effects of $\lambda_{R1}$ and $\lambda_{R2}$ are negligible, the energy spectrum is derived as

$$\xi = s_z M \pm \sqrt{\eta^2 v_F^2 k^2 + (lE_z - \eta s_z \lambda_{SO})^2}. \tag{3}$$

The effect of $E_z$ is to change the mass of the Dirac electron. When increasing $E_z$ from $E_z = 0$ with $M$ fixed, the mass decreases (increases) for the Dirac cone characterized by $\eta s_z = +1$ ($\eta s_z = -1$) until $E_z = \lambda_{SO}/l$, but the behavior becomes opposite after $E_z = \lambda_{SO}/l$. As a result, the tip of each Dirac cone is pushed either downward or upward as indicated in Fig.12. Consequently the valley symmetry is broken. Conducting VPM phase occupies a primary part of the phase diagram, where electrons are removed from the $K$ to $K'$ valley when $E_z > 0$ and $M > 0$. As a result, $E_f$ is below the top of the valence band of the $K$ valley but above the bottom of the conduction band of the $K'$ valley although a band gap is opened at the two valleys. Marginal VPM (M-VPM) states appear on the phase boundaries, where the bottom of the conduction band of the $K'$ valley and the top of the valence band of the $K$ valley touch the Fermi level.

Strong magnetic moments can be induced by doping 3$d$ transition metals. Silicene decorated with certain 3$d$ transition metals (such as vanadium) has been demonstrated to sustain a stable QAHE using both analytical model and first principles theory.[42] Electrically tunable topological states can also be realized in certain other transition metal doped silicene. Silicene decorated with chemical functional groups (X-Si, X= -H, -F, -Cl,-Br, and -I) are also investigated.[43] They are trivial band insulators with larger band gaps. By applying biaxial tensile strain, X-Si can be transformed into QSH state. These findings are highly desirable for future nanoelectronic and spintronic applications.

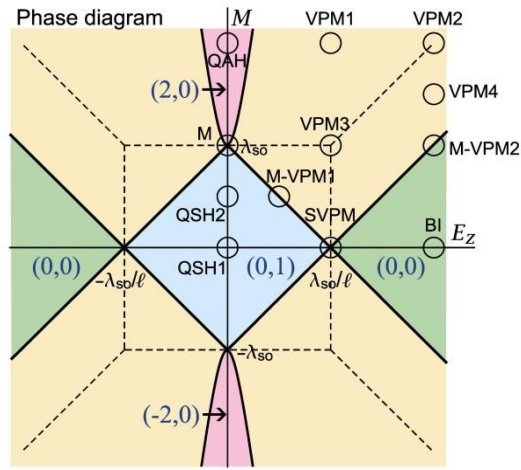

**Figure 11.** Phase diagram in the $E_z$-$M$ plane. Heavy lines represent phase boundaries, where the system becomes metallic. Chern and spin-Chern numbers ($C$, $C_s$) are well defined and given in insulator phases. Dotted lines represent the points where the band gap closes, which are within the VPM phase. A circle shows a point where the energy spectrum is calculated and shown in Fig. 12. Reproduced with permission from Ref. [38]. Copyright 2012 American Physical Society.

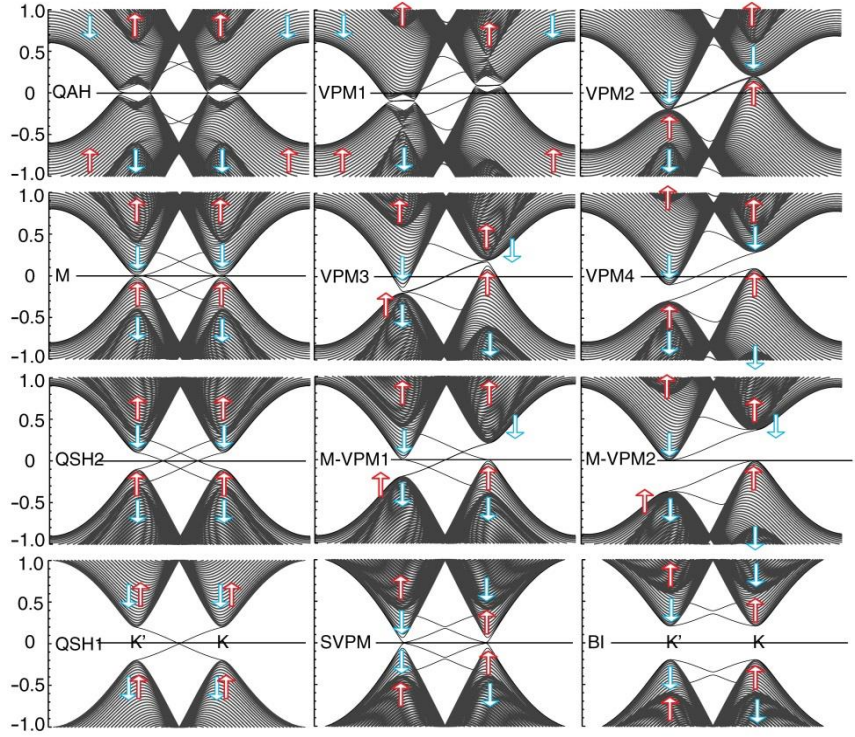

**Figure 12.** The band structure of a silicene nanoribbon at marked points in the phase diagram (Fig. 11). The vertical axis is the energy in unit of *t*, and the horizontal axis is the momentum. We can clearly see the Dirac cones representing the energy spectrum of the bulk. Lines connecting the two Dirac cones are edge modes. The spin $s_z$ is practically a good quantum number, which we have assigned to the Dirac cones. Reproduced with permission from Ref. [38]. Copyright 2012 American Physical Society.

In addition to the quantum states described above, broken inversion symmetry induced by $E_z$ could arouse quantum valley Hall (QVH) effect in silicene, where Dirac fermions in different valleys flow to the opposite transverse edges with quantized conductivity when an in-plane electric field is applied. When $E_f$ lies in the gap, the spin and valley Hall conductivities are obtained as[41]

$$\sigma_{xy}(\text{spin}) = [\sigma_{xy}(K,\uparrow) - \sigma_{xy}(K,\downarrow)] + [\sigma_{xy}(K',\uparrow) - \sigma_{xy}(K',\downarrow)]$$
$$= -\frac{e^2}{2h}[\text{sgn}(\lambda_{SO} - lE_z) + \text{sgn}(\lambda_{SO} + lE_z)] \quad (4)$$

$$\sigma_{xy}(\text{valley}) = [\sigma_{xy}(K,\uparrow) + \sigma_{xy}(K,\downarrow)] - [\sigma_{xy}(K',\uparrow) + \sigma_{xy}(K',\downarrow)]$$
$$= \frac{e^2}{2h}[\text{sgn}(\lambda_{SO} - lE_z) - \text{sgn}(\lambda_{SO} + lE_z)]. \quad (5)$$

When $|E_z| < |E_c| = \dfrac{\lambda_{SO}}{l}$ then $\sigma_{xy}(\text{spin}) = -\dfrac{e^2}{h}$ and $\sigma_{xy}(\text{valley}) = 0$, which corresponds to the QSH state as shown and discussed in Figs. 11 and 12. When $|E_z| > |E_c| = \dfrac{\lambda_{SO}}{l}$ then $\sigma_{xy}(\text{spin}) = 0$ and $\sigma_{xy}(\text{valley}) = \dfrac{e^2}{h}$, which corresponds to a QVH insulating state. As shown in Fig. 13, when $E_f$ is in the band gap, the QSH (blue) and QVH (red) conductivities show a sharp transition as electric field changes. When $E_f$ is in the conduction or valence band, this topological transition is also possible to be observed as shown in Fig. 13 (green and magenta). As the SOC in silicene is strong, the Hall plateau won't be affected at finite low temperature.

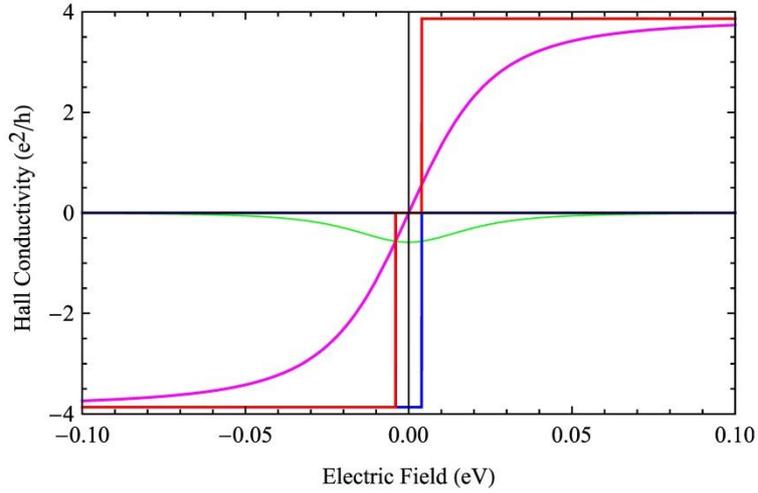

**Figure 13.** Quantum spin and valley Hall conductivities as a function of the perpendicular electric field for a fixed value of the SOI (3.9 meV). Blue and green: Quantum spin Hall conductivities for the Fermi level in the band gap and the conduction band, respectively. Red and magenta: Valley Hall conductivities for the Fermi level in the band gap and the conduction band, respectively. Reproduced with permission from Ref. [41]. Copyright 2013 AIP Publishing LLC.

A new quantum state of matter-the valley-polarized QAH state is also predicted in silicene by Yao *et. al.*.[44] When the time-reversal symmetry is broken from the exchange field, the competition between intrinsic and extrinsic Rashba SOCs results in a new topological phase. When the extrinsic Rashba SOC gradually increases from $\lambda_{R1}/t = 0.01$ to $\lambda_{R1}/t = 0.09$, as shown in Fig. 14, the bulk band gap at valley $K'$ closes and reopens twice. Through analyzing the resulting Berry curvatures of the occupied valence bands, the nonzero Chern number ($C = -1$) directly indicates a QAH phase as shown in Fig. 14d. Surprisingly, the two valleys contribute to different Chern numbers, *i.e.*, $C_K = 1$ but $C_{K'} = -2$. This imbalance signals a QVH phase with valley-Chern number $C_v = 3$, which means the proposed state is a valley-polarized QAH state. As shown in Fig. 15, there are three edge states localized at each boundary. For the upper (lower) boundary, two edge states associated with valley $K'$ propagate from right (left) to left (right) while one associated with valley $K$ counterpropagates from left (right) to right (left), leading to a valley-polarized edge current.

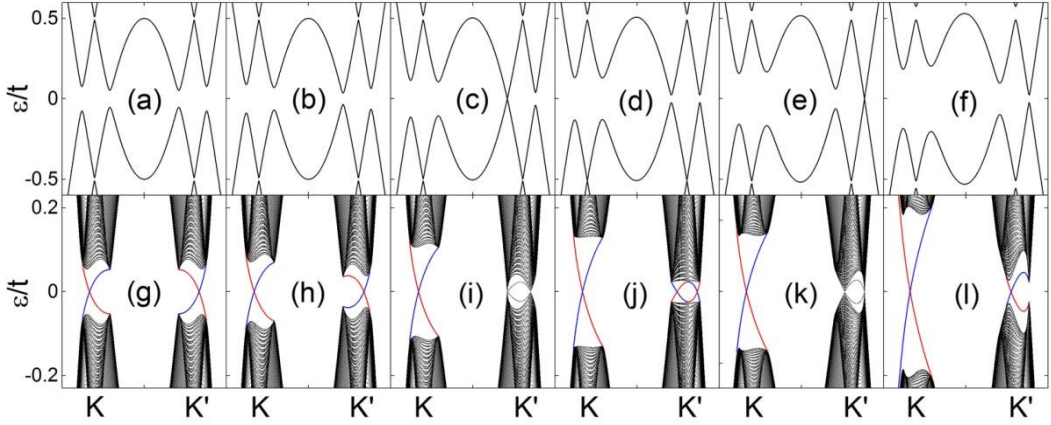

**Figure 14.** Evolution of band structures of the bulk [(a)–(f)] and zigzag-terminated [(g)–(l)] silicene as a function of extrinsic Rashba SOC $\lambda_{R1}$ at fixed intrinsic Rashba SOC $\lambda_{R2}$ and exchange field *M*. (a) $\lambda_{R1} = 0$. Bulk energy gaps open around the *K* and *K'* Dirac points. The size of the bulk gap near valley *K* is exactly the same as that near valley *K'*. (b)–(f) $\lambda_{R1}/t =$ 0.01, 0.031, 0.045, 0.067, 0.09, respectively. Along with the increasing of $\lambda_{R1}$, the bulk gap around valley *K* gradually increases, while the bulk gap near valley *K'* closes twice [see panels (c) and (e)] and reopens twice [see panels (d) and (f)]. (g)–(l) The valley-associated gapless edge modes at valley *K* is unchanged, but those for valley *K'* change; *i.e.*, there are two or one pair of edge modes after the bulk gap reopens. Colors are used to label the edge modes localized at opposite boundaries. Other parameters in Eq. 2 are set to be $\lambda_{SO}/(3\sqrt{3}t)$ =0.002, $2\lambda_{R2}/(3t)$ =0.08, and $M/t$ = 0.5. Reproduced with permission from Ref. [44]. Copyright 2014 American Physical Society.

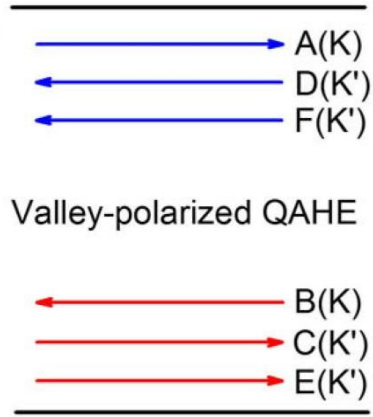

**Figure 15.** Valley-associated edge modes for the valley-polarized QAHE. Colors are used to label the edge modes localized at opposite boundaries. Reproduced with permission from Ref. [44]. Copyright 2014 American Physical Society.

A spin-filter can also be designed based on a quantum point contact (QPC) in a silicene sheet characterized by a short and narrow constriction as shown in Fig. 16a.[39] To make the valley degrees of freedom well separated, the zigzag edges are adopted for the whole geometry along the direction of current flow. The two opposite wide regions are in the SVPM phase with $E_f > 0$ to model metallic source and drain (Fig. 16c) and the gated constriction is in the M-VPM phase with a fixed $E_z > E_c$ and an applied Zeeman field $M = lE_z - \lambda_{SO}$ (Fig. 16d). An electrostatic potential barrier $U(x_i)$ is added along the current flow direction as shown in Fig. 16a, which is non-vanishing only in the constriction region. The resulting spin polarization as a function of the effective chemical potential, $\mu_0 = E_f - U_0$, is shown in Fig. 16b. Through quantum transport calculations based on the iterative Green's function method, the QPC produces an almost fully spin-polarized current. The spin-polarization direction can be easily reversed by locally changing the potential barrier via gating control in the constriction. Such a high-efficiency, field-tunable spin-filter based on silicene takes advantage of charge carriers in the bulk system, small Rashba SOC, and controllable spin-splitting due to IS breaking, and is robust against weak disorder (compared with $\lambda_{SO}+M$) and edge imperfections.

A Y-shaped spin/valley separator based on silicene shown in Fig. 17 is proposed to separate the two spin/valley polarizations from the incoming lead 1, with one flowing to lead 2 and the other flowing to lead 3. An out-of-plane electric field, $E_z > E_c$, is first applied in the central silicene sheet to create a nonvanishing Berry curvature, and chemical potential $\mu$ is tuned into the conduction bands. Then by setting potentials, for example, $V_1 > V_2 = V_3$, at the terminals of silicene, charge carriers acquire an anomalous velocity proportional to the Berry curvature in the transverse direction, similar to that reported by Xiao *et al.* in graphene.[45] By using linear response theory with negligible $\lambda_{R1}$ and $\lambda_{R2}$, the spin and valley Hall conductivity for $E_f$ in the conduction band are obtained

$$\sigma_{xy}(\text{spin}) = -\frac{e^2}{h}\left(\frac{\lambda_{SO}}{\mu}\right) \quad (6)$$

$$\sigma_{xy}(\text{valley}) = \frac{e^2}{h}\left(1 - \frac{lE_z}{\mu}\right). \quad (7)$$

Thus valley and hence spin polarization imbalance at output terminals $V_2$ and $V_3$ are obtained (with opposite polarization between them). Therefore silicene provides an ideal platform for efficiently manipulating spin/valley degrees of freedom.

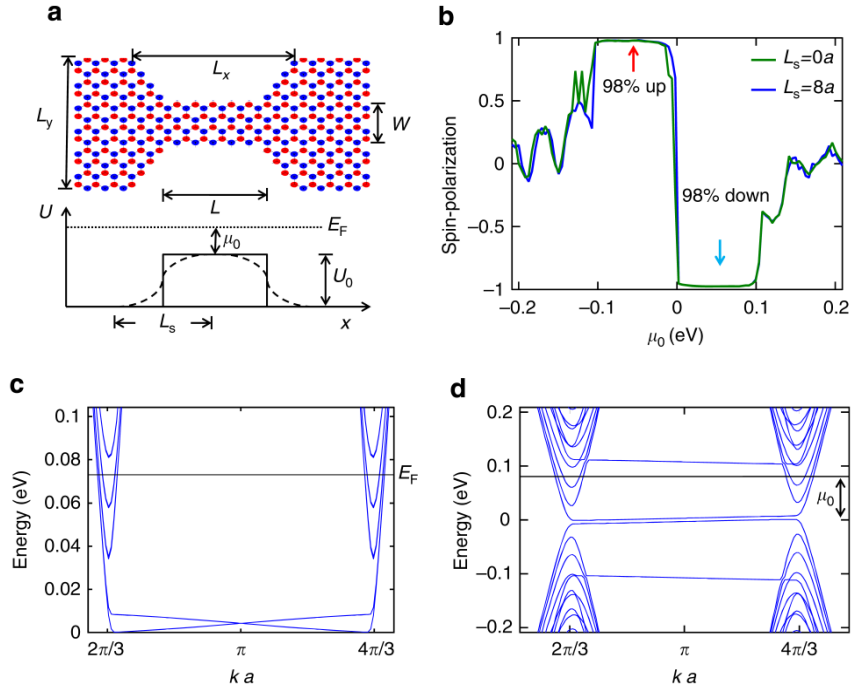

**Figure 16. Tunable high-efficiency spin-filter.** (a) Geometry of the spin-filter and the profile of the potential barrier $U(x_i)$. The two colored atoms on the lattice emphasize the buckled structure. (b) Spin-polarization of the filter as a function of $\mu_0$ in the constriction. The blue (green) line corresponds to the case of applying potential barrier with a rectangular (smooth) shape. (c,d) are typical dispersion relations for the wide and the constriction regions,

respectively. Reproduced with permission from Ref. [39]. Copyright 2013 Macmillan Publishers Limited.

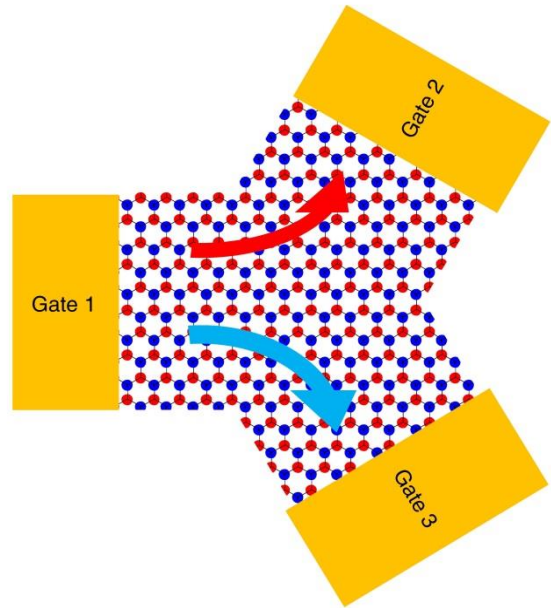

**Figure 17. Y-shaped silicene spin-separator.** A schematic Y-shape separator of silicene with three local gates (leads). The current separates into the gate 2 and gate 3, respectively, carrying opposite spin/valley degrees of freedom. Reproduced with permission from Ref. [39]. Copyright 2013 Macmillan Publishers Limited.

## Conclusion and outlook

In conclusion, tremendous progress has been made in silicene spintronic study over the past several years. Zigzag silicene nanoribbons become half-metals by a transverse electric filed or asymmetric edge modifications, and spin-filter and spin FET based on them are designed with high performance. Giant magnetoresistance is obtained in zigzag silicene nanoribbons either by utilizing the switch between different magnetic states or the symmetry-dependent transport property. Half-metallicity can also be achieved in semihydrogenated silicene by applying an out of plane gate voltage. The spin-filters designed by utilizing half-metallicity in semihydrogenated silicene and by utilizing the peculiar topological quantum states in pristine silicene are predicted to be able to deliver nearly 100% spin-polarized current. By adjusting an electric field and/or an exchange field, quantum spin Hall effect, quantum anomalous Hall effects, valley-polarized quantum anomalous Hall effects, and quantum valley Hall effects can be expected in pristine silicene. However, so far no experimental observations have been reported despite all these fascinating theoretical predictions. Because the extreme surface sensitivity greatly limits both the isolation and device fabrications of silicene, any experimental fundamental and device research on silicene becomes extremely difficult. Even though the existed difficulties, very recently a silicene FET operating at room temperature has been reported via a growth-transfer-fabrication process.[23] This experimental breakthrough is expected to strongly stimulate researchers in the field to continue looking for methods to solve the problems concerned and making further improvement. We believe that silicene spintronics will come true one day as more mature experimental techniques are developed.

## ACKNOWLEDGMENT

This work was supported by the National Natural Science Foundation of China (No. 11274016/11474012), the National Basic Research Program of China (No. 2013CB932604/ 2012CB619304), and the National Science Foundation Grant (1207141).

# References

[1] Du X, Skachko I, Barker A and Andrei E Y 2008 *Nature Nanotech.* **3** 491-495

[2] Morozov S V, Novoselov K S, Katsnelson M I, Schedin F, Elias D C, Jaszczak J A and Geim A K 2008 *Phys. Rev. Lett.* **100** 016602

[3] Tombros N, Jozsa C, Popinciuc M, Jonkman H T and van Wees B J 2007 *Nature* **448** 571-574

[4] Han W, Pi K, McCreary K M, Li Y, Wong J J I, Swartz A G and Kawakami R K 2010 *Phys. Rev. Lett.* **105** 167202

[5] Yang T Y, Balakrishnan J, Volmer F, Avsar A, Jaiswal M, Samm J, Ali S R, Pachoud A, Zeng M, Popinciuc M, Güntherodt G, Beschoten B and Özyilmaz B 2011 *Phys. Rev. Lett.* **107** 047206

[6] Pesin D and MacDonald A H 2012 *Nature Mater.* **11** 409-416

[7] Han W, Kawakami R K, Gmitra M and Fabian J 2014 *Nature Nanotech.* **9** 794-807

[8] Lin C-C, Penumatcha A V, Gao Y, Diep V Q, Appenzeller J and Chen Z 2013 *Nano Lett.* **13** 5177-5181

[9] Wang W H, Pi K, Li Y, Chiang Y F, Wei P, Shi J and Kawakami R K 2008 *Phys. Rev. B* **77** 020402

[10] Kim W Y and Kim K S 2008 *Nature Nanotech.* **3** 408-412

[11] Son Y-W, Cohen M L and Louie S G 2006 *Nature* **444** 347-349

[12] Ozaki T, Nishio K, Weng H and Kino H 2010 *Phys. Rev. B* **81** 075422

[13] Lakshmi S, Roche S and Cuniberti G 2009 *Phys. Rev. B* **80** 193404

[14] Kang J, Wu F and Li J 2011 *Appl. Phys. Lett.* **98** 083109

[15] Feng B, Ding Z, Meng S, Yao Y, He X, Cheng P, Chen L and Wu K 2012 *Nano Lett.* **12** 3507-3511

[16] Chen L, Liu C-C, Feng B, He X, Cheng P, Ding Z, Meng S, Yao Y and Wu K 2012 *Phys. Rev. Lett.* **109** 056804

[17] Chen L, Li H, Feng B, Ding Z, Qiu J, Cheng P, Wu K and Meng S 2013 *Phys. Rev. Lett.* **110** 085504

[18] Vogt P, De Padova P, Quaresima C, Avila J, Frantzeskakis E, Asensio M C, Resta A, Ealet B and Le Lay G 2012 *Phys. Rev. Lett.* **108** 155501

[19] Chiappe D, Grazianetti C, Tallarida G, Fanciulli M and Molle A 2012 *Adv. Mater.* **24** 5088-5093

[20] Fleurence A, Friedlein R, Ozaki T, Kawai H, Wang Y and Yamada-Takamura Y 2012 *Phys. Rev. Lett.* **108** 245501

[21] Meng L, Wang Y, Zhang L, Du S, Wu R, Li L, Zhang Y, Li G, Zhou H, Hofer W A and Gao H-J 2013 *Nano Lett.* **13** 685-690

[22] Chiappe D, Scalise E, Cinquanta E, Grazianetti C, van den Broek B, Fanciulli M, Houssa M and Molle A 2013 *Adv. Mater.* **26** 2096-2101

[23] Tao L, Cinquanta E, Chiappe D, Grazianetti C, Fanciulli M, Dubey M, Molle A and Akinwande D 2015 *Nature Nanotech.* **10** 227-231

[24] Appelbaum I, Huang B and Monsma D J 2007 *Nature* **447** 295-298

[25] Huang B, Monsma D J and Appelbaum I 2007 *Phys. Rev. Lett.* **99** 177209

[26] Huang B, Jang H-J and Appelbaum I 2008 *Appl. Phys. Lett.* **93** 162508

[27] Sanvito S 2011 *Chem. Soc. Rev.* **40** 3336-3355

[28] Liu C-C, Feng W and Yao Y 2011 *Phys. Rev. Lett.* **107** 076802

[29] Yao Y, Ye F, Qi X-L, Zhang S-C and Fang Z 2007 *Phys. Rev. B* **75** 041401

[30] Wang Y, Zheng J, Ni Z, Fei R, Liu Q, Quhe R, Xu C, Zhou J, Gao Z and Lu J 2012 *Nano* **07** 1250037

[31]   Ding Y and Wang Y 2013 *Appl. Phys. Lett.* **102** 143115

[32]   Yang X F, Liu Y S, Feng J F, Wang X F, Zhang C W and Chi F 2014 *J. Appl. Phys.* **116** 124312

[33]   Zhang D, Long M, Zhang X, Cao C, Xu H, Li M and Chan K 2014 *Chem. Phys. Lett.* **616–617** 178-183

[34]   Deng X, Zhang Z, Tang G, Fan Z, Zhu H and Yang C 2014 *Sci. Rep.* **4** 4038(doi: 10.1038/srep04038)

[35]   Xu C, Luo G, Liu Q, Zheng J, Zhang Z, Nagase S, Gao Z and Lu J 2012 *Nanoscale* **4** 3111

[36]   Kang J, Wu F and Li J 2012 *Appl. Phys. Lett.* **100** 233122

[37]   Pan F, Quhe R, Ge Q, Zheng J, Ni Z, Wang Y, Gao Z, Wang L and Lu J 2014 *Physica E* **56** 43-47

[38]   Ezawa M 2012 *Phys. Rev. Lett.* **109** 055502

[39]   Tsai W-F, Huang C-Y, Chang T-R, Lin H, Jeng H-T and Bansil A 2013 *Nat Commun* **4** 1500

[40]   Tahir M and Schwingenschlögl U 2013 *Sci. Rep.* **3** 1075 (doi:10.1038/srep01075)

[41]   Tahir M, Manchon A, Sabeeh K and Schwingenschlögl U 2013 *Appl. Phys. Lett.* **102** 162412

[42]   Zhang X-L, Liu L-F and Liu W-M 2013 *Sci. Rep.* **3** 2908 (doi:10.1038/srep02908)

[43]   Cao G, Zhang Y and Cao J 2015 *Phys. Lett. A* **379** 1475-1479

[44]   Pan H, Li Z, Liu C-C, Zhu G, Qiao Z and Yao Y 2014 *Phys. Rev. Lett.* **112** 106802

[45]   Xiao D, Yao W and Niu Q 2007 *Phys. Rev. Lett.* **99** 236809